\documentclass[runningheads]{llncs}
\usepackage{graphicx}
\graphicspath{{figures/}}  % GR edit. Paths can be of the form {./figs/} or {subdir1/}{subdir2/}{subdir3/}...{subdirn/}
\usepackage{epstopdf} % GR edit
\usepackage{mathtools} 										% split environment, \vcentcolon, \coloneqq for define equals
\renewcommand{\v}[1]{\ensuremath{\mathbf{#1}}} 				% bold vectors except Greek letters \v{u} 
\newcommand{\uv}[1]{\ensuremath{\mathbf{\hat{#1}}}} 		% unit vectors in bold \uv{u}, \uv{e}_z,...
\usepackage[binary-units=true]{siunitx} 	% e.g. \SI{2.86e-6}{\micro\meter} or \si{\milli\second\per\meter\squared} or \num{1e4}
\usepackage{placeins} 						% for \FloatBarrier
\newcommand{\bpar}[0]{\ensuremath{b_{\parallel}}}
\newcommand{\bperp}[0]{\ensuremath{b_{\perp}}}
\newcommand{\bperphalf}[0]{\ensuremath{\frac{b_{\perp}}{2}}}
\newcommand{\lampar}[0]{\ensuremath{\lambda_{\parallel}}}
\newcommand{\lamperp}[0]{\ensuremath{\lambda_{\perp}}}
\newcommand{\Ssing}[0]{\ensuremath{S_{\textrm{sing}}}}
\newcommand{\Scros}[0]{\ensuremath{S_{\textrm{cros}}}}
\newcommand{\Sinplane}[0]{\ensuremath{S_{\textrm{ip}}}}
\newcommand{\eccenD}[0]{\ensuremath{\varepsilon_D}}
\newcommand{\peakindex}[0]{SPSI}
\newcommand{\peakindexm}[0]{\ensuremath{\textrm{\peakindex{}}}}
\newcommand{\clmthird}[0]{\ensuremath{\left(c_L-\frac{1}{3}\right)}}
\newcommand{\third}[0]{\ensuremath{\frac{1}{3}}}
\newcommand{\expbrace}[1]{\ensuremath{\exp\left\{#1\right\}}}

\usepackage{amsfonts}
\usepackage{amssymb}
\usepackage{subfig}	% for \subfloat (replaces the deprecated subfigure package and \subfigure command)

\begin{document}
\title{A Signal Peak Separation Index for axisymmetric B-tensor encoding}
%
%\titlerunning{Abbreviated paper title}
% If the paper title is too long for the running head, you can set
% an abbreviated paper title here
%
\author{Ga\"{e}tan Rensonnet\inst{1,2} %\orcidID{0000-0002-3988-462X}
\and Jonathan Rafael-Pati{\~n}o\inst{1}
\and Beno\^{i}t Macq \inst{2}
\and Jean-Philippe Thiran \inst{1,3,4}
\and Gabriel Girard$^*$ \inst{1,3,4}
\and Marco Pizzolato$^*$ \inst{5,1}
}
\authorrunning{G. Rensonnet et al.}
% First names are abbreviated in the running head.
% If there are more than two authors, 'et al.' is used.
%
\institute{Signal Processing Lab (LTS5), \'{E}cole polytechnique f\'{e}d\'{e}rale de Lausanne, Lausanne, Switzerland
%\email{gaetan.rensonnet@epfl.ch}
\and ICTEAM Institute, Universit\'{e} catholique de Louvain, Louvain-la-Neuve, Belgium
\and Center for BioMedical Imaging (CIBM),  Lausanne, Switzerland
\and Radiology Department, Centre hospitalier universitaire vaudois and University of Lausanne, Lausanne, Switzerland
\and Department of Applied Mathematics and Computer Science, Technical University of Denmark, Kongens Lyngby, Denmark
}
\def\thefootnote{}\footnotetext{Preprint submitted to the MICCAI 2020 International Workshop on Computational Diffusion MRI (CDMRI'20).}
\def\thefootnote{*}\footnotetext{These senior authors contributed equally.}
\def\thefootnote{\arabic{footnote}}

\maketitle              % typeset the header of the contribution
%

%\textit{Postprint accepted at the MICCAI 2020 International Workshop on Computational Diffusion MRI (CDMRI'20). The final paper will be published in the Computational Diffusion MRI Book as part of the Springer Visualization in Mathematics series. An appendix containing more detailed mathematical derivations was included with a arXiv preprint (\url{http://doi_link}).}

\begin{abstract}
Diffusion-weighted MRI (DW-MRI) has recently seen a rising interest in planar, spherical and general B-tensor encodings. Some of these sequences have aided traditional linear encoding in the estimation of white matter microstructural features, generally by making DW-MRI \emph{less} sensitive to the orientation of axon fascicles in a voxel. However, less is known about their potential to make the signal \emph{more} sensitive to fascicle orientation, especially in crossing-fascicle voxels. 
Although planar encoding has been commended for the resemblance of its signal with the voxel's orientation distribution function (ODF), linear encoding remains the near undisputed method of choice for orientation estimation. This paper presents a theoretical framework to gauge the sensitivity of axisymmetric B-tensors to fascicle orientations. A signal peak separation index (\peakindex{}) is proposed, motivated by theoretical considerations on a simple multi-tensor model of fascicle crossing. Theory and simulations confirm the intuition that linear encoding, because it maximizes B-tensor anisotropy, possesses an intrinsic advantage over all other axisymmetric B-tensors. At identical \peakindex{} however, oblate B-tensors yield higher signal and may be more robust to acquisition noise than their prolate counterparts. The proposed index relates the properties of the B-tensor to those of the tissue microstructure in a straightforward way and can thus guide the design of diffusion sequences for improved orientation estimation and tractography.

\end{abstract}
\section{Introduction}
Diffusion-weighted magnetic resonance imaging (DW-MRI) is based on the application of time-varying external magnetic-field gradients $\v{g}(t)$ to probe water diffusion~\cite{stejskal1965spin}. In the brain white matter, it is mainly used for two purposes. One is a necessary first step for tractography\index{tractography} consisting in estimating the orientation of the main fascicles of axons in a voxel, often via the orientation distribution function (ODF)\index{orientation distribution function}, and is referred to as \emph{orientation estimation}\index{orientation estimation}. Another is to estimate finer microstructural properties of those fascicles such as the morphology of their axons, generally referred to as \emph{microstructure estimation}~\cite{basser2002diffusion}.

In both tasks, the focus has traditionally been on \emph{linear encoding}\index{linear encoding} in which $\v{g}(t)\in\mathbb{R}^3$ is parallel to a fixed direction $\uv{u}\in \mathbb{S}^2$ for all $t$, the best-known example of which being the pulsed-gradient spin-echo (PGSE)~\cite{stejskal1965spin}. More recently, there has been growing interest in more general gradient waveforms living in a 2D plane, known as \emph{planar encoding}~\cite{wedeen2012geometric,ozarslan2015rotating}\index{planar encoding}, or the whole 3D space, referred to as general multidimensionnal or \emph{B-tensor encoding}~\cite{mori1995diffusion,drobnjak2011optimising,westin2014measurement,westin2016q}\index{b-tensor encoding}. As the name indicates, such sequences are often studied through their associated symmetric, positive-definite B-tensor defined as $\v{B}\coloneqq \gamma^2\int_0^T\int_0^t\int_0^t \v{g}(t_1)\cdot \v{g}(t_2)^\top \,\mathrm{d}t_1\,\mathrm{d}t_2\;\mathrm{d}t \in \mathbb{R}^{3\times3}$, where $\gamma$ is the gyromagnetic ratio of protons and $T$ the duration of the sequence. B-tensors with 1, 2 and 3 identical, strictly positive eigenvalues refer to linear, planar and spherical encoding respectively~\cite{westin2014measurement,westin2016q}\index{spherical encoding}.

Those general waveforms have been mostly used for microstructure estimation~\cite{mori1995diffusion,jespersen2013orientationally,lasivc2014microanisotropy}, especially to resolve degeneracies wherein different microstructural properties are difficult to estimate simultaneously using conventional linear encoding only; e.g., 
extracting axonal microstructural properties irrespective of their orientation 
with 3D B-tensor encoding~\cite{topgaard2019diffusion,avram2019measuring}; 
disentangling volume or signal fractions and diffusivities with 
	(planar) double diffusion encoding (DDE)~\cite{coelho2019resolving} or 
	(spherical) triple diffusion encoding~\cite{jensen2018characterizing}; 
separating microscopic anisotropy from orientation dispersion using DDE~\cite{lawrenz2013double,jespersen2013orientationally} or spherical encoding~\cite{lasivc2014microanisotropy,szczepankiewicz2015quantification,cottaar2020improved}.

How well these general waveforms may perform at orientation estimation is still an open question. Linear encoding maximizes B-tensor anisotropy~\cite{eriksson2015nmr} and is thus expected to provide high signal sensitivity to the orientation of anisotropic structures. Spherical encoding on the other hand minimizes B-tensor anisotropy~\cite{lasivc2014microanisotropy,eriksson2015nmr,szczepankiewicz2015quantification,jensen2018characterizing,topgaard2019diffusion,avram2019measuring,cottaar2020improved} and probably offers little benefit. Planar encoding DW-MRI data directly reflects the ODF of the voxel without needing the post-processing or modeling typically required in linear encoding~\cite{wedeen2012geometric,ozarslan2015rotating}. It has been shown to compare to~\cite{wedeen2012geometric} and potentially outperform~\cite{ozarslan2015rotating} linear encoding.

This paper proposes a framework to assess the potential of waveforms characterized by an axisymmetric B-tensor for providing DW-MRI data suited to orientation estimation. A signal peak separation index (\peakindex{})\index{signal peak separation index} is proposed, motivated by theoretical considerations on a toy model of fascicle crossing\index{crossing fascicles} assuming a superposition of diffusion tensors. The index relates the properties of the B-tensors to microstructural properties such as the crossing angle, the NMR-apparent volume fraction and the microscopic anisotropy\index{microscopic anisotropy} of the fascicles, to quantify the directional information content of the signal. Theoretical predictions are made about the respective merits of linear, planar and intermediate encodings which are then verified in simulation experiments.

\section{Theory}
In this work, a diffusion sequence is entirely characterized by an axisymmetric tensor $\v{B}$ with eigenvalues $\left\{\frac{\bperp}{2}, \frac{\bperp}{2}, \bpar\right\}$. The orientation $\uv{u}_B$ of $\v{B}$ is defined as the eigenvector associated with $\bpar$, which is not necessarily the largest eigenvalue, and the b-value, a measure of diffusion weighting, as $b\!\coloneqq \!\textrm{tr}\left(\v{B}\right)\!=\!\bperp\!+\!\bpar$. 
%Diffusion sequence entirely characterized by a symmetric, positive definite B tensor assumed to be axisymmetric, i.e. the details of the magnetic field gradient waveforms used to obtain B are abstracted away. This B is characterized by a parallel direction uB associated with an eigenvalue bpar (note that parallel preferred over main or principal because it is not necessarily the largest eigenvalue) and two identical eigenvalues bperp associated with orthogonal directions perpendicular to uB: B=R(B)B0R(B)T where D0=diag(bperp/2, bperp/2, bpar) and R(B) is the 3D orthonormal rotation matrix verifying R(B)TR(B) = R(B)TR(B) = I3 and serving as a change of coordinates operator ez=R(u)u, i.e. it transforms the reference laboratory coordinates into coordinates in its the frame.
The linearity coefficient $c_L$ is defined as $c_L\! \coloneqq \! \frac{\bpar}{b}$ and characterizes B-tensor encodings as planar ($c_L\!=\!0$), planar-like or oblate ($0\!< \! c_L \! < \! \third{}$), spherical ($c_L\!=\!\third{}$), linear-like or prolate ($\third{}\!<c_L<\!1$) and linear ($c_L\!=\!1$). It is related to the tensor anisotropy $b_{\Delta}$~\cite{eriksson2015nmr} via $b_{\Delta}=\frac{3}{2}\clmthird{}\in \left[-\frac{1}{2},1\right]$. In this setting, the exact temporal profiles of the physically-applied magnetic-field gradients are thus ignored.

The diffusion of water within a fascicle of bundled axons is represented by an axisymmetric diffusion tensor $\v{D}$, referred to as ``zeppelin'', with eigenvalues $\left\{\lamperp, \lamperp, \lampar\right\}$, where $\lampar>\lamperp$ is enforced, and with orientation $\uv{u}_D$ defined as its principal eigenvector. The normalized DW-MRI signal $\Ssing$ arising from a fascicle characterized by a zeppelin $\v{D}$ subject to $\v{B}$ is $\exp\left(-\v{B}:\v{D}\right)$~\cite{neeman1990pulsed}, where $:$ denotes the Frobenius inner product. As detailed in Appendix~\ref{app:single_zeppelin_signal}, this yields
\begin{equation}
\begin{split}
\Ssing & \left(\v{B};\v{D}\right) = \\ 
 & \exp\left(-\bperphalf\lamperp 
		  - \cos^2\left(\varphi_D-\varphi_B\right) \left(\bperphalf\lamperp + \bpar\lampar\right) 
		  - \sin^2\left(\varphi_D-\varphi_B\right) \left(\bpar\lamperp + \bperphalf\lampar \right)
   \right),
\end{split}
\label{eq:sig_sing}	  
\end{equation}
showing that the signal is effectively a function of the angle $\left|\varphi_D-\varphi_B\right|$ between $\uv{u}_D$ and $\uv{u}_B$,
with $\varphi_D$ and $\varphi_B$ their azimuthal coordinates in their common plane, defined from an arbitrary common reference. 
In the spherical case $c_L\!=\!\frac{1}{3}$, all values of $\varphi_B$ identically lead to $\Ssing=\exp\left(-\frac{b}{3}\left(2\lamperp+\lampar\right)\right)$.

\subsection{A toy model of fascicle crossing under B-tensor encoding}
The voxel-level signal $\Scros$ resulting from the crossing of two populations characterized by $\v{D}_1$ and $\v{D}_2$ with NMR-apparent volume fractions, referred to as signal fractions, $\nu_1$ and $\nu_2=1-\nu_1$ under the diffusion-encoding tensor $\v{B}$ is approximated by the following superposition~\cite{rensonnet2018assessing}
\begin{equation*}
\Scros \left(\v{B}; \v{D}_1, \v{D}_2\right) = \nu_1 \Ssing  \left(\v{B};\v{D}_1\right) + \nu_2 \Ssing  \left(\v{B};\v{D}_2\right).
%\label{eq:superposition}
\end{equation*}
Our toy model assumes identical microstructural properties ($\v{D}_1$ and $\v{D}_2$ have identical eigenvalues) and Fascicle 1 as the dominant fascicle ($\nu_1 \geq \nu_2$).

The in-plane signal $\Sinplane$ of a crossing is defined as the signal for $\uv{u}_B$ lying in the plane spanned by $\uv{u}_{1}$ and $\uv{u}_2$, assumed non collinear. It is the only signal contribution relevant to peak detection because the out-of-plane signal, defined when $\uv{u}_B \cdot \uv{u}_1 \!= \!\uv{u}_B\cdot\uv{u}_2 \!=\!0$, is equal to $\exp\left(-\bperphalf\left(\lampar+\lamperp\right)-\bpar\lamperp\right)$ and therefore holds no information about the fascicles' orientations $\uv{u}_{1}$ and $\uv{u}_2$. Setting without loss of generality $\varphi_1\!=\!0$, $\varphi_2 \!=\! \alpha$ with $\alpha$ the crossing angle between $\uv{u}_1$ and $\uv{u}_2$, and noting $\varphi_B$ the in-plane azimuthal coordinate of $\uv{u}_B$, $\Sinplane$ is computed as
\begin{equation}
\begin{split}
\Sinplane & \left(\varphi_B\right)  =  \nu_1 \Ssing\left(\varphi_B\right) + \nu_2 \Ssing\left(\alpha-\varphi_B\right)  \\
    &=  \nu_1 \exp\left(- \bperphalf\lamperp 
					- \cos^2\left(\varphi_B\right) \left(\bperphalf\lamperp + \bpar\lampar\right) 
					- \sin^2\left(\varphi_B\right) \left(\bpar\lamperp + \bperphalf\lampar\right) \right) \\
    & + \nu_2 \exp\left(- \bperphalf\lamperp 
						- \cos^2\left(\alpha\!-\!\varphi_B\right) \left(\bperphalf\lamperp + \bpar\lampar\right)
						- \sin^2\left(\alpha\!-\!\varphi_B\right) \left(\bpar\lamperp + \bperphalf\lampar\right)  \right),
\end{split}
\label{eq:sig_inplane}
\end{equation}
where the single argument to $\Ssing$ refers to the angle between $\uv{u}_D$ and $\uv{u}_B$, with the dependence on all other properties of $\v{B}$ and $\v{D}$ implied. 

\begin{figure}
\centering
\includegraphics[scale=0.41]{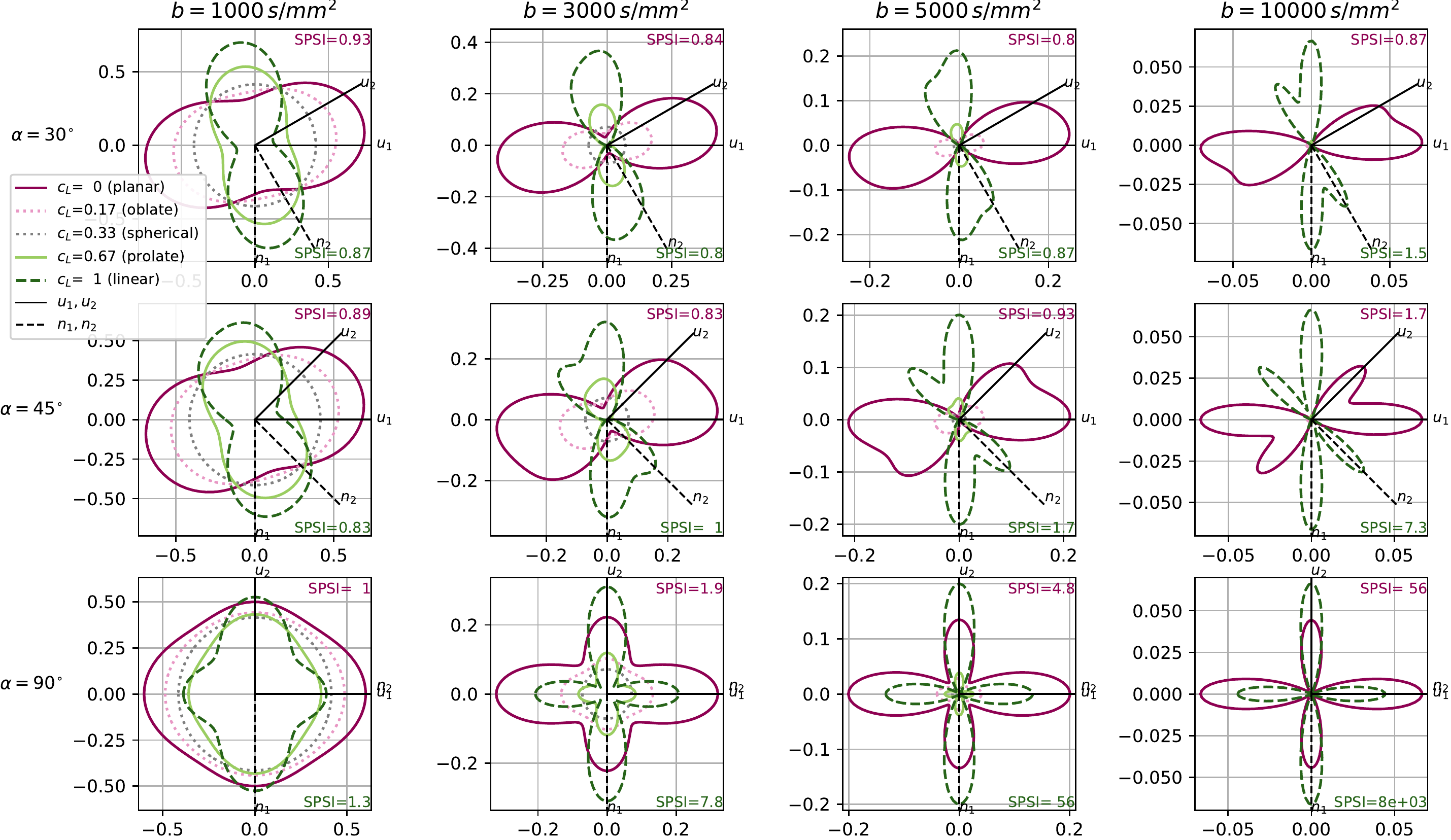}
\caption{\textbf{Linear encoding is more sensitive to fascicle orientation than planar encoding while intermediate cases offer little benefit.} The peaks and troughs in the in-plane signal $\Sinplane$ of our toy model, especially those associated with the smaller peak (here $\nu_2=0.4$), only become visible at sufficiently high $b$, large $\alpha$ and extremal value of $c_L$, which is captured by values of our proposed \peakindex{} staying below or exceeding 1.}
%Spherical encoding displays a constant circle and reveals no directional information.
% iso: constant circle, combine disadvantages of both methods. Continuous black lines depict the orientationss $\uv{u}_1$ and $\uv{u}_2$ of the two zeppelins representing the diffusion in fascicles. The dashed black lines show the directions $\uv{n}_1$ and $\uv{n}_2$ normal to those orientations.
\label{fig:sig_inplane}
\end{figure}

Radial plots of $\Sinplane$ are shown in Fig.~\ref{fig:sig_inplane} for various values of $b, \alpha$ and $c_L$ and display the characteristic butterfly shape of DW-MRI signals. Signals from planar-like ($c_L\!\! <\!\!\third{}$) and linear-like ($c_L\!\!>\!\!\third{}$) encodings are $90^{\circ}$ out of phase, with high signal obtained respectively in the range $\varphi_B\in\left[0,\alpha\right]$ and $\varphi_B \!\in\! \left[-\frac{\pi}{2},\alpha-\frac{\pi}{2}\right]$. The signal increases when magnetic-field gradients are applied perpendicular to $\uv{u}_1$ and $\uv{u}_2$, ``against'' the fascicles (see Fig.~\ref{fig:planar_vs_linear}).

Mathematically, voxel-level fascicles are distinguishable when the high-signal range of 
$\Sinplane$ exhibits two distinct maxima separated by one minimum. Intuitively, 
the maxima of $\Sinplane\left(\varphi_B\right)$ are expected to be around $0$ and 
$\alpha$ (corresponding to $\uv{u}_1$ and $\uv{u}_2$)
 in planar-like encoding, 
and around $\!-\!\frac{\pi}{2}$ and $\alpha\!-\!\frac{\pi}{2}$ (corresponding to the directions $\uv{n}_1$ and $\uv{n}_2$ normal to $\uv{u}_1$ and $\uv{u}_2$) 
in linear-like encoding. 
Likewise, the minima of $\Sinplane\left(\varphi_B\right)$ are 
expected to lie about halfway between the maxima, along the bisector $\uv{b}_u$ for 
planar- and $\uv{b}_n$ for linear-like encoding. 
In planar-like encoding for instance, if $\Sinplane$ has a lower value along $\uv{u}_2$ than along
 $\uv{b}_u$, it suggests that the second peak in the high-signal range of $\Sinplane$ is too
  small or hasn't appeared yet (see Fig.~\ref{fig:sig_inplane} at $b=\SI{3000}{\second\per\milli\meter\squared}$ 
  and $\alpha=45^{\circ}$) and that the signal does not contain the orientational information required for accurate ODF or fascicle orientation estimation. The small fascicle ($\nu_2\!<\!\nu_1$) is only detectable when its associated signal maximum
(if present at all) exceeds the signal dip (if present at all) between the fascicles, as apparent in Fig.~\ref{fig:sig_inplane}. These considerations motivate the definition of a signal index based on (approximate) peaks and troughs of the signal, as presented in the following section.

\subsection{The signal peak separation index}
The signal peak separation index (\peakindex) of an axisymmetric B-tensor encoding is defined as the ratio of the signal ``against'' the smaller fascicle to the signal acquired along the bisector of the fascicles in a toy model of identical intersecting zeppelins (see Fig.~\ref{fig:planar_vs_linear}) and reads
\begin{equation}
\peakindexm \coloneqq 
\left\{
\begin{array}{ll}
\frac{\Sinplane\left(\alpha\right)}{\Sinplane\left(\alpha/2\right)} & \hspace{5mm} \textrm{for } c_L \leq \third{} \\
\frac{\Sinplane\left(\alpha-\pi/2\right)}{\Sinplane\left(\alpha/2-\pi/2\right)} & \hspace{5mm} \textrm{for } c_L > \third{}.
\end{array}
\right.
\label{eq:def_index_via_ratio}
\end{equation}

Using Eq.~\eqref{eq:sig_inplane}, Eq.~\eqref{eq:def_index_via_ratio} becomes (see Appendix~\ref{app:max_to_min_ratio})
\begin{equation}
\begin{split}
\peakindexm \left(c_L;b, \alpha, \nu_1, \eccenD{}\right) & = 
 \nu_1 \exp\left(-\frac{3}{2} \sin\left(\frac{\alpha}{2}\right) \sin\left(\frac{3\alpha}{2}\right) \left|c_L-\frac{1}{3}\right| b \eccenD{} \right) \\
 	& \hspace{5mm}+ \nu_2 \exp\left( \frac{3}{2} \sin^2\left(\frac{\alpha}{2}\right)  \left|c_L-\frac{1}{3}\right| b \eccenD{} \right),
\end{split}
\label{eq:def_index}
\end{equation}
where $\eccenD{}\coloneqq\left(\lampar - \lamperp\right)$ represents the microscopic anisotropy of each fascicle. The numerator in Eq.~\eqref{eq:def_index_via_ratio} is in general not equal, mathematically, to the true maximum of the in-plane signal $\Sinplane$ associated with $\uv{u}_2$ and therefore underestimates the value of the signal peak. Similarly, the denominator is in general an overestimation of the true trough of $\Sinplane$ between the fascicles. The proposed \peakindex{} is therefore a conservative underestimate of the true peak-to-trough ratio, which makes $\peakindexm{} > 1$ \emph{a sufficient but not always necessary condition for signal peaks to be separated}, in noiseless settings. As discussed in more details below, the true peaks and troughs of $\Sinplane$, when they exist, are actually found along directions fairly close to those selected in Eq.~\eqref{eq:def_index_via_ratio}.
When $\peakindexm{} > 1$, higher \peakindex{} should indicate better orientation estimation performance; when $\peakindexm{} < 1$, the fascicles are often indistinguishable in the signal and changes in \peakindex{} values become less interpretable.\\

\subsubsection{Linear encoding maximizes signal contrast.}
As shown in Fig.~\ref{fig:ratio_vs_cL}, $\peakindexm{}\left(c_L\right)$ is symmetric around $c_L\!=\!\third{}$ and strictly convex on either side of $c_L\!=\!\third{}$. It can thus only be maximized at the boundaries of the subintervals, i.e., $c_L\!=\!0,\third{},\textrm{ or }1$. % TODO: cite convex optimization book like Boyd's ?
Assuming $\peakindexm(0)>1$ leads to $\peakindexm(\frac{2}{3})=\peakindexm(0)>1$, by symmetry. 
 Based on the strict convexity for $c_L\in\left[\third{}, 1\right]$ and because $\peakindexm(\third{})=1$, $\peakindexm{}(c_L)$ must then be increasing, not decreasing, for $c_L\geq \frac{2}{3}$. 
 Therefore, $\peakindexm(1)\!>\!\peakindexm(\frac{2}{3})=\peakindexm(0)$ holds, i.e., \emph{in all cases where $ \exists c_L \textrm{s.t. }\peakindexm{}\left(c_L\right) \! >\! 1$, linear encoding always achieves a higher \peakindex{} than planar encoding}. Since maximizing $\left|c_L-\third{} \right| $ maximizes \peakindex{}, intermediate oblate and prolate encodings are sub-optimal.\\
% A corollary is that, in ranges of $c_L$ where $\peakindexm{}\left(c_L\right)>1$, fully planar (linear) encoding always achieves a higher \peakindex{} than any intermediate oblate (prolate) encoding.\\
 % at a given b-value, for zeppelins with a given anisotropy $\eccenD$ intersecting at a given angle $\alpha$.\\

\subsubsection{Planar encoding provides higher signal.}
At equal \peakindex{}, Eq.~\eqref{eq:sig_inplane} reveals that planar-like encoding yields higher signal than linear-like encoding with
\begin{equation}
\frac{\Sinplane{}\left(\varphi_B;c_L=\third{}\!-\!\Delta_L \right)}{\Sinplane{}\left(\varphi_B\!-\!\frac{\pi}{2};c_L=\third{}\!+\!\Delta_L \right)} = \exp\left(\frac{b}{2}\Delta_L \eccenD{} \right) > 1 \quad \forall \varphi_B,
\label{eq:ratio_planar_over_linear_same_anisotropy}
\end{equation}
valid for any $\alpha, \nu_1$, for $0 < \Delta_L\leq \third{}$ (the ``deviation'' from spherical), which is reminiscent of the link between $\eccenD{}$ and a ratio of linear and spherical signal derived in~\cite{cottaar2020improved}. Appendix~\ref{app:all_ratios} provides additional variations of similar ratios. The signal from fully planar ($c_L\!=\!0$) encoding turns out to also exceed that of fully linear ($c_L\!=\!1$) encoding $\forall \varphi_B$, for $b\eccenD{}>0$ and $\nu_2>0$
\begin{equation}
\begin{split}
&\frac{\Sinplane{}\left(\varphi_B;c_L\!=\!0 \right)}{\Sinplane{}\left(\varphi_B\!-\!\frac{\pi}{2};c_L\!=\!1 \right)} \\
 &= \exp\left(\sin^2\left(\varphi_B\right) \frac{b}{2}\eccenD{} \right)   \cdot
 \left[\frac{\nu_1 + \nu_2 \exp\left( \sin\left(\alpha\right) \sin\left(2\varphi_B\!-\!\alpha\right)\frac{b}{2}\eccenD{} \right) }
            {\nu_1 + \nu_2 \exp\left( \sin\left(\alpha\right) \sin\left(2\varphi_B\!-\!\alpha\right)   b       \eccenD{} \right)}
 \right] > 1,
\end{split}
\label{eq:ratio_planar_over_linear_pure}
\end{equation}
the proof of which is presented in Appendix~\ref{app:proof_difficult_ratio}.

\subsection{Theoretical support for \peakindex{}}
The true locations of the extrema of $\Sinplane\left(\varphi_B\right)$ are roots of its first derivative $\frac{\partial\Sinplane}{\partial \varphi_B}$, i.e. solutions of the non-linear equation
\begin{equation}
\begin{split}
\nu_2 \sin\left(2\varphi_B \right) &= \\
 \nu_1 \sin\left(2\left(\alpha-\varphi_B\right)\right) & \exp\left(\frac{3}{2}  \sin\left(\alpha\right) \sin\left(\alpha-2\varphi_B\right) \clmthird{} b \eccenD{} \right)\!\!.
\end{split}
\label{eq:zeros_derivative}
\end{equation}
As hinted at in the previous section and shown by the circles in Fig.~\ref{fig:sig_vs_phiB}, Eq.~\eqref{eq:zeros_derivative} is in general \emph{not} satisfied by the values $\alpha, \frac{\alpha}{2}, \alpha\!-\!\frac{\pi}{2}$ and $\frac{\alpha\!-\!\pi}{2}$ used in our definition of \peakindex{} (Eq.~\eqref{eq:def_index_via_ratio}). 
%This is visible in Fig.~\ref{fig:sig_vs_phiB} where the values of $\varphi_B$ used for \peakindex{}, indicated by circles, are slightly off the true locations of the peaks and troughs of $\Sinplane$. 
In this section, two particular cases are studied in which Eq.~\eqref{eq:zeros_derivative} admits closed-form solutions. Whether these solutions are minima or maxima of $\Sinplane$ is then determined by the sign of the second derivative $\frac{\partial^2\Sinplane}{\partial \varphi_B^2}$.

\subsubsection{Equal fascicle contributions.} 
The first special case is $\nu_1\!=\!\nu_2\!=\!0.5$. Equation~\eqref{eq:zeros_derivative} is solved by $\varphi_B\!=\!\frac{\alpha}{2}$ and $\varphi_B\!=\!\frac{\alpha\!-\!\pi}{2}$, yielding candidate extrema along the bisectors $\uv{b}_u$ and $\uv{b}_n$. The second derivative is computed as
\begin{subequations}
\begin{eqnarray*}
&\left. \frac{\partial^2\Sinplane}{\partial \varphi_B^2}  \right|_{\varphi_B=\frac{\alpha}{2}} 
      = \frac{3}{2} \Ssing\left(\frac{\alpha}{2}\right) \clmthird{} b \eccenD{} \left[\frac{3}{2} \clmthird{} b \eccenD{} \sin^2\left(\alpha\right) + 2\cos\left(\alpha\right) \right] \nonumber \\
&\left. \frac{\partial^2\Sinplane}{\partial \varphi_B^2}  \right|_{\varphi_B=\frac{\alpha-\pi}{2}}
       = \frac{3}{2} \Ssing\left(\frac{\alpha-\pi}{2}\right) \clmthird{} b \eccenD{} \left[\frac{3}{2} \clmthird{} b \eccenD{} \sin^2\left(\alpha\right) - 2\cos\left(\alpha\right) \right], \nonumber 
\end{eqnarray*}
\end{subequations}
which actually holds for any value of $\nu_1$. This shows that $\uv{b}_u$ in planar-like ($c_L\! <\! \third{}$) and $\uv{b}_n$ in linear-like ($c_L \!>\! \third{}$) encoding are minima, not maxima, when
\begin{equation}
\frac{3}{2}\left|c_L-\frac{1}{3}\right| b \eccenD{} > \frac{2\cos\left(\alpha\right)}{\sin^2\left(\alpha\right)}.
\label{eq:equal_nus_condition_biss}
\end{equation}
Equation~\eqref{eq:equal_nus_condition_biss} is satisfied for large enough $b$, microscopic anisotropy $\eccenD{}$, crossing angle $\alpha$ and anisotropy $\left|c_L-\frac{1}{3}\right|$, giving linear encoding (with $c_L\!=\!1$) an advantage over planar encoding (with $c_L\!=\!0$) as it can reach a value double that of planar encoding ($\frac{2}{3}$ vs $\frac{1}{3}$). 
It can reasonably be assumed that these conclusions extend to the general case $\nu_1\neq\nu_2$ 
unless $\nu_1\gg 0.5$ and that, in general, the signal troughs occur near the locations of the bisectors when Eq.~\eqref{eq:equal_nus_condition_biss} holds, thus justifying the denominator of Eq.~\eqref{eq:def_index_via_ratio} defining \peakindex{}.

\subsubsection{Right-angle crossing.}
The second special case is $\alpha=\frac{\pi}{2}$. Using Eq.~\eqref{eq:zeros_derivative}, the roots of $\frac{\partial\Sinplane}{\partial \varphi_B}$ are found to be
\begin{subequations}
\begin{eqnarray}
& \sin\left(2\varphi_B\right)=0 \Leftrightarrow \varphi_B = k \pi /2\; \left(k\in \mathbb{Z}\right), \label{eq:zeros_1}\\
\textrm{or } & \cos\left(2\varphi_B\right) = \frac{2\log\left(\nu_1/\nu_2\right)}{3 \clmthird{} b \eccenD{}}. \label{eq:zeros_2}
\end{eqnarray}
\end{subequations}

Equation~\eqref{eq:zeros_1} indicates candidate extrema along ($\varphi_B\!=-\frac{\pi}{2},0$) and perpendicular ($\varphi_B\!=0,\frac{\pi}{2}$) to the fascicles. The second derivative of $\Sinplane$ 
\begin{subequations}
\begin{eqnarray*}
&\left. \frac{\partial^2\Sinplane}{\partial \varphi_B^2}  \right|  _{\varphi_B=0}
   = 3 \clmthird{} b \eccenD{} \left(\nu_1 \Ssing\left(0\right) -\nu_2 \Ssing\left(\frac{\pi}{2}\right)  \right) \\
&\left. \frac{\partial^2\Sinplane}{\partial \varphi_B^2}  \right|  _{\varphi_B=\pm\frac{\pi}{2}} 
   = 3 \clmthird{} b \eccenD{} \left(\nu_2 \Ssing\left(0\right) -\nu_1 \Ssing\left(\frac{\pi}{2}\right)  \right) 
\end{eqnarray*}
\end{subequations}
reveals, after using Eq.~\eqref{eq:sig_sing}, that acquisitions with physical gradients against the larger fascicle ($\varphi_B\!=\!\alpha$ in planar-like and $\varphi_B\!=\!\alpha\!-\!\frac{\pi}{2}$ in linear-like encoding, $\alpha\!=\!\frac{\pi}{2}$) always correspond to signal maxima ($\frac{\partial^2\Sinplane}{\partial \varphi_B^2} < 0$) and that acquisitions against the smaller fascicle are also signal maxima whenever
\begin{equation}
\log\left(\frac{\nu_1}{\nu_2}\right) < \frac{3}{2}\left|c_L-\frac{1}{3}\right| b \eccenD{},
\label{eq:right_angle_cond_maxima}
\end{equation}
which again occurs with sufficiently large $b$ and $\eccenD{}$, $c_L$ far enough from $\third{}$ and $\nu_1$ close enough to $\nu_2$. Equation~\eqref{eq:zeros_1} and \eqref{eq:right_angle_cond_maxima} likely describe general trends extending to the case $\alpha\leq \frac{\pi}{2}$, justifying the numerator in Eq.~\eqref{eq:def_index_via_ratio} defining \peakindex{}.\\

Condition~\eqref{eq:right_angle_cond_maxima} also guarantees that the cosine in Eq.~\eqref{eq:zeros_2} takes a value in the feasible range $\left[-1,1\right]$. Solving Eq.~\eqref{eq:zeros_2} then leads to candidate extrema at
\begin{equation}
\varphi_B=\pm \frac{1}{2} \arccos\left( \frac{2\log\left(\nu_1/\nu_2\right)}{3 \clmthird{} b \eccenD{}} \right),
\label{eq:right_angle_min_loc}
\end{equation}
corresponding to locations between the fascicles. Those extrema are minima of $\Sinplane$ when $\frac{\partial^2\Sinplane}{\partial \varphi_B^2} \!>\! 0$, which is ensured by the sufficient (possibly too restrictive) condition
\begin{equation}
\frac{9}{4} \left(c_L-\frac{1}{3}\right)^2 b^2 \eccenD{}^2 > 1,
\label{eq:right_angle_cond_minima}
\end{equation}
easily attained in practice. Equation~\eqref{eq:right_angle_min_loc} further reveals that those signal troughs are obtained with $\uv{u}_B$ close to the bisectors $\uv{b}_u$ and $\uv{b}_n$, in line with the conclusions of the previous particular case $\nu_1=\nu_2=0.5$ and justifying the denominator of Eq.~\eqref{eq:def_index_via_ratio}. 
For a relatively extreme case $\nu_1=0.8$, with $\frac{3}{2}\clmthird{}b=\SI{3000}{\second\per\milli\meter\squared}$ and $\eccenD{}=\SI{2}{\micro\meter\squared\per\second}$, the signal troughs are located at $\left|\varphi_B\right|= 38.3^{\circ}$ instead of $45^{\circ}$, i.e. a difference of only $14.8\%$.

In general, a small error on the location of an extremum of a continuously differentiable function leads to a limited error on the signal value since the derivative is almost zero, and the function thus more or less flat, in the vicinity of an extremum.

\section{Methods}

\subsection{Verification of theoretical predictions}
The proposed \peakindex{} was computed as a function of $c_L$, for various values of $b$ and $\alpha$ with $\nu_1\!=\!0.6$ to verify that \peakindex{} was systematically maximized, in so far as it ever reached the threshold value of 1, by linear rather than planar encoding. 
The in-plane signal $\Sinplane$ was then computed for various values of $b$, $\alpha$, $c_L$ with $\nu_1\!=\!0.6$ to visually verify whether $\Sinplane\left(\alpha\right)$ and
 $\Sinplane\left(\alpha\!-\!\pi/2\right)$ were close to true signal peaks and  $\Sinplane\left(\alpha/2\right)$ and $\Sinplane\left(\alpha/2\!-\!\pi/2\right)$ close to true signal troughs. The corresponding \peakindex{} was computed for each scenario to assess that i) $\peakindexm{}\!<\!1$ was associated to in-plane signals that did not exhibit two clear, separate peaks; ii) in regimes where $\peakindexm{} \!>\! 1$, higher \peakindex{} was linked to sharper separation of signal peaks; iii) at fixed \peakindex{} and $\left|c_L\!-\!\third{}\right|\!<\!\third{}$, signal for $c_L\!<\!\third{}$ was higher than for $c_L\!>\!\third{}$.\\

\subsection{Robustness of \peakindex{} beyond our toy model}
The goal of this experiment was to i) verify that \peakindex{} correlated with accuracy of fascicle orientation estimation even when the assumptions of the toy model, which motivated its definition, were not met; ii) evaluate different types of B-tensor encoding at orientation estimation in the presence of acquisition noise.

In order to account for an intra- and extra-axonal compartment, each fascicle was modeled by a ``stick'' ($\lamperp\!=\!0, \lampar\!=\!\SI{2.2}{\micro\meter\squared\per\milli\second}$) aligned with a zeppelin ($\lamperp\!=\!\SI{0.4}{\micro\meter\squared\per\milli\second}, \lampar\!=\!\SI{1.5}{\micro\meter\squared\per\milli\second}$) with lower parallel $\lampar$ following evidence on small animals~\cite{kunz2018intra}, with intra-fascicle signal fractions $0.65$ and $0.35$ respectively. Values of $\nu_1\!=\!0.6$ and $\alpha\!=\!60^{\circ}$ were used. The signal was simulated for B-tensors with $b=\SI{3000}{\second\per\milli\meter\squared}$, varying $c_L\!\in\!\left[0,1\right]$ and 200 orientations $\uv{u}_B$ uniformly distributed on the sphere. %, generated using an electrostatic repulsion algorithm. 
The signal was corrupted by Rician noise with signal-to-noise ratio (SNR) defined for all $c_L$ as $\textrm{SNR}=\frac{1}{\sigma_g}$, with $\sigma_g$ the standard deviation of the Gaussian noise process in the receiver coils. For each value of SNR and $c_L$, 90 independent crossing-fascicle voxels were simulated.

The estimation of fascicle orientation was performed with DIPY~\cite{garyfallidis2014dipy} routines by first reconstructing the ODF from the noisy signal and then extracting the maxima from the ODF. Two strategies were considered to reconstruct the ODF. In the first one, for planar-like encoding ($c_L\!\!<\!\!1/3$), following~\cite{ozarslan2015rotating}, the ODF was directly computed as a spherical harmonics (SH) fit of the DW-MRI signal with maximum degree $l=10$ and a Laplace-Beltrami regularization factor $\lambda=0.001$~\cite{descoteaux2007regularized} while for linear-like encoding ($c_L\!\!>\!\!1/3$) the constant solid angle (CSA) model~\cite{aganj2010reconstruction} was used. In the second strategy, constrained spherical deconvolution (CSD)~\cite{tournier2007robust} was identically applied to all encodings. The single-fascicle response function required by CSD was recursively calibrated in a data-driven way from an initial rotational harmonics (RH) fit of the signal from a ``fat'' zeppelin~\cite{tax2014recursive} (FA=0.20, trace=$\SI{2.2}{\micro\meter\squared\per\milli\second}$) for each specific B-tensor type. To perform that calibration automatically, 10 single-fascicle voxels using the \emph{same} stick-and-zeppelin fascicle model were included. In theory, the algorithm could thus filter out the noise and estimate the ideal single-fascicle response function. In both cases the ODF was computed on 724 points uniformly spread on the unit sphere. An ODF value was considered a peak if it exceeded the ODF minimum by at least $15\%$ of the total ODF range. A minimum angular separation of $15^{\circ}$ between detected maxima was enforced and only the 3 maxima with largest ODF values were kept. The angular error $\theta$ was computed by iterating over the true orientations $\uv{u}_k$ ($k=1,\dots,M_{\textrm{true}}$) as~\cite{canales2019sparse}
\begin{equation}
\theta = \frac{1}{M_{\textrm{true}}}\sum_{k=1}^{M_{\textrm{true}}}\min_m\left\{\arccos\left(\left|\uv{e}_m\cdot \uv{u}_k\right|\right)\right\},
\end{equation}
where $\uv{e}_m$ is the unit orientation of the $m$-th detected peak and $M_{\textrm{true}}=2$ the true number of fascicles. The mean angular error was then computed as the mean $\mathbb{E}\left[\theta\right]$ over all noise repetitions for each type of encoding at each SNR level. Finally, \peakindex{} was computed for each tested value of $c_L$ using the groundtruth values for $b, \nu_1$ and $\alpha$ but setting $\eccenD{}$, which is not defined in the case of a multi-compartment fascicle model, to a generic value of $\SI{1.8}{\micro\meter\squared\per\milli\second}$.

\section{Results}
\subsection{Verification of theoretical predictions}
\begin{figure}
\centering
% METRIC DIAGRAM
\subfloat[Schematic definition of \peakindex{}.]
{\includegraphics[width=3.1cm, height=4cm, clip=true, trim=0cm 0cm 23.75cm 6.25cm]{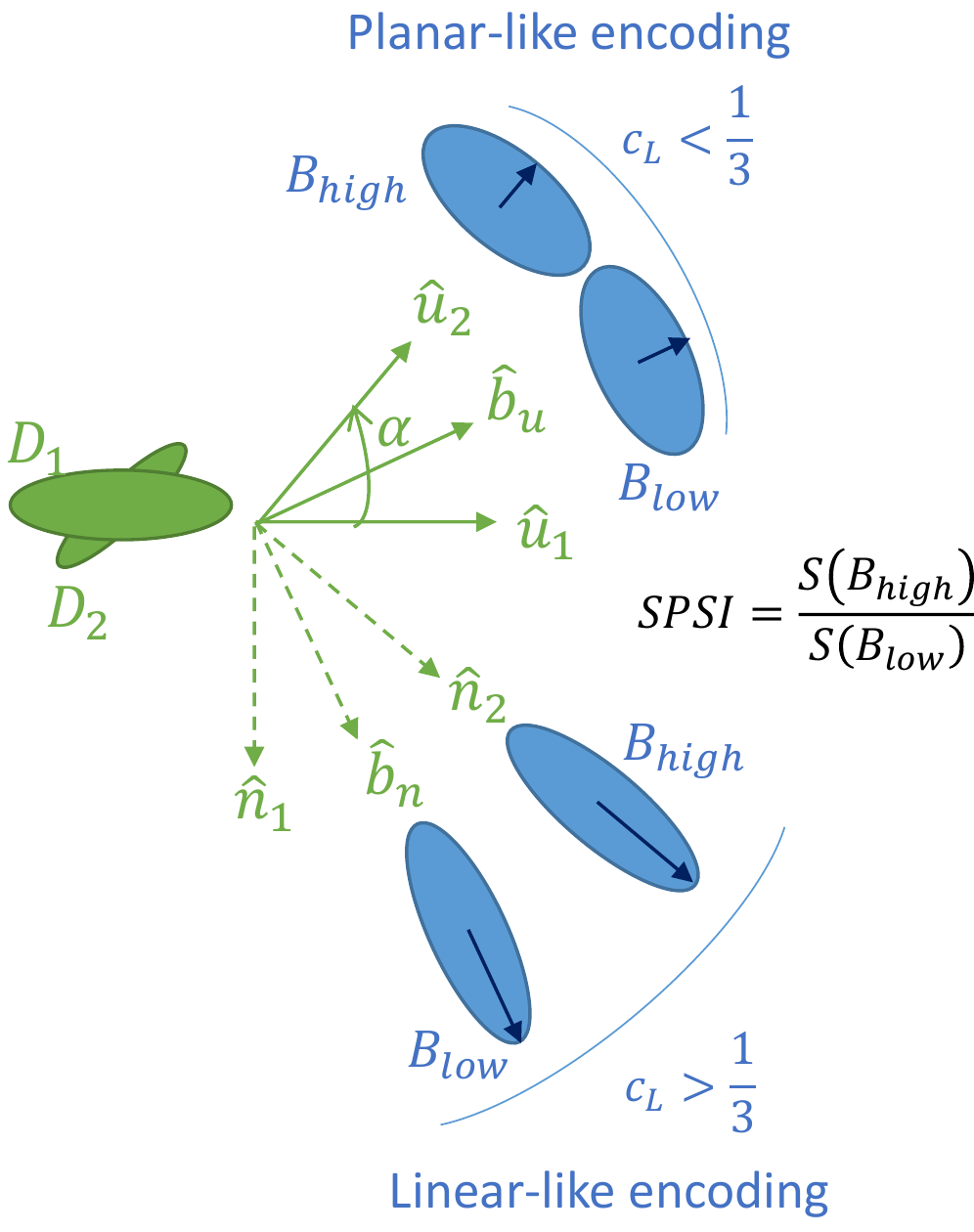}
 \label{fig:planar_vs_linear}}
 \hspace{0.2cm}
% CONVEXITY OF index(C_L)
\subfloat[\peakindex{} vs shape of the B-tensor $c_L$ with $\nu_1=0.6$. The y-axis was trimmed for clarity at $b=\num{5000}$ and $b=\num{10000}$.]
{\includegraphics[width=8.5cm, height=4.0cm]{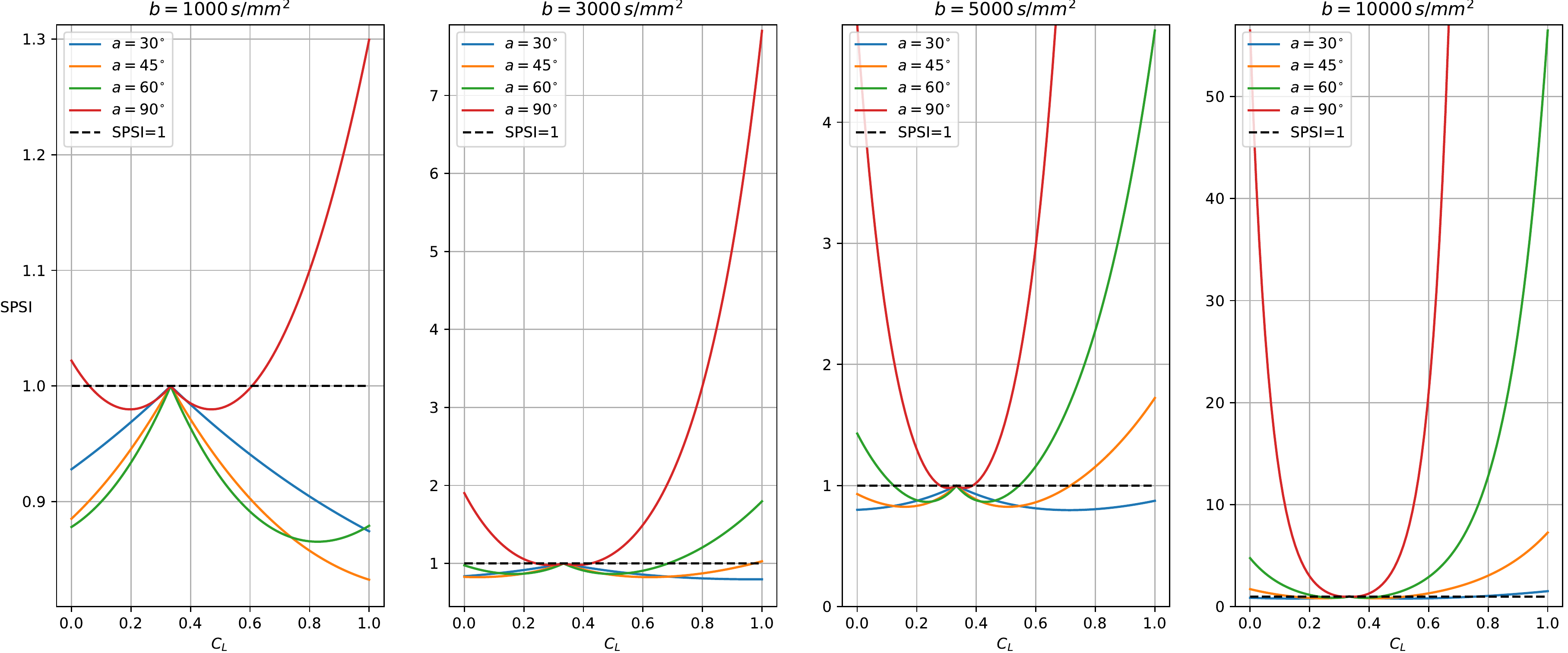}
\label{fig:ratio_vs_cL}}
\caption{\textbf{The proposed signal peak separation index (\peakindex{}) is a ratio of signal against the smaller fascicle to signal between the fascicle. It is maximized by linear, rather than planar, encoding.}}
\end{figure}

\begin{figure}
\centering
\includegraphics[width=\textwidth]{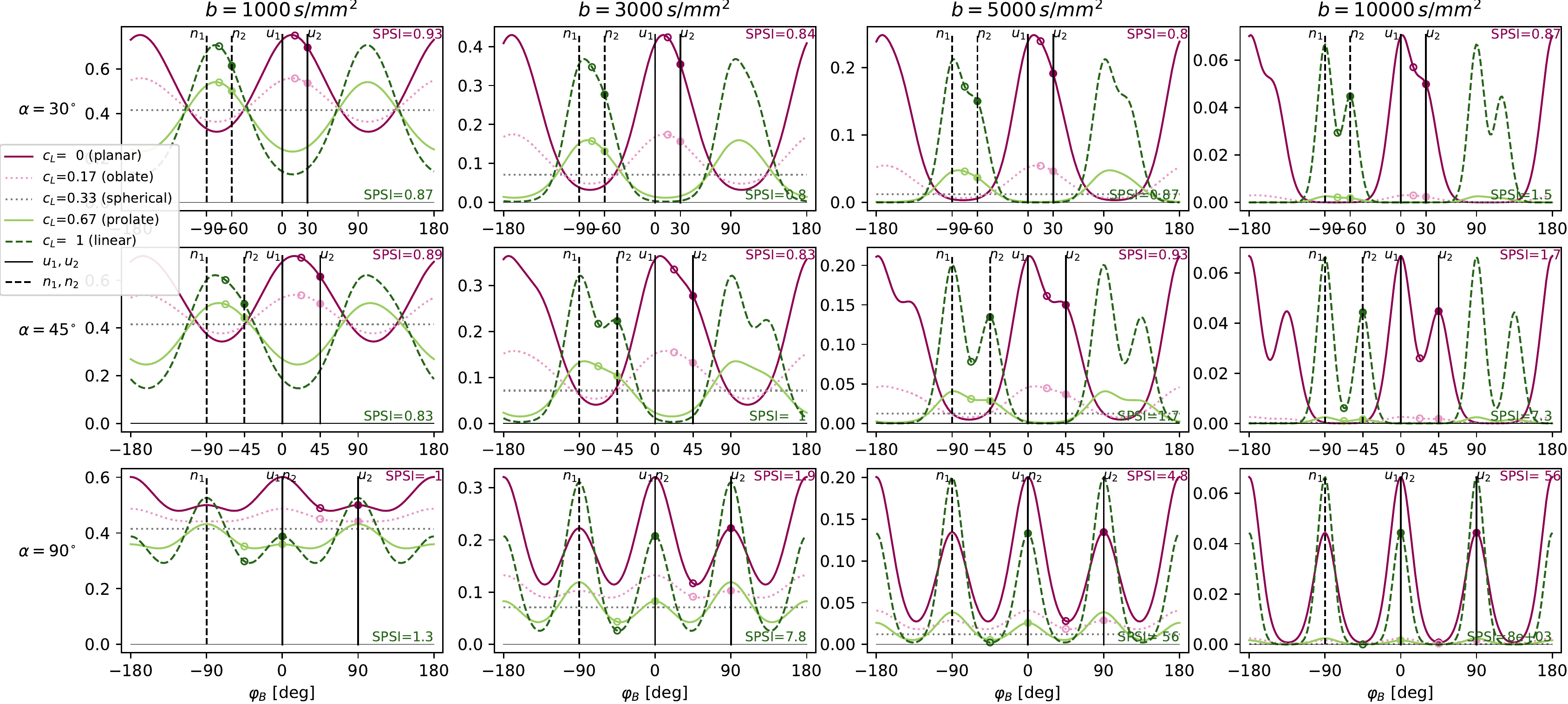}
\caption{\textbf{Our proposed (\peakindex{}) accurately approximates the ratio of the true signal peaks and troughs.}
The in-plane signal $\Sinplane$ acquired ``against'' the smaller fascicle (see filled circles) and along the main or normal bisector (empty circles), used to define \peakindex{} in Eq.~\eqref{eq:def_index_via_ratio}, seem very close to the true extrema, under the hypotheses of our toy model ($\nu_1=0.6$ here).
}
\label{fig:sig_vs_phiB}
\end{figure}

In Fig.~\ref{fig:ratio_vs_cL}, all \peakindex{} curves crossing the $\peakindexm{}=1$ threshold were systematically maximized by linear encoding at $c_L=1$, as predicted by theory. \peakindex{} was particularly sensitive to $\alpha$ and $b$. 
Figure~\ref{fig:sig_vs_phiB} confirms that the signal values used in Eq.~\eqref{eq:def_index_via_ratio} for the \peakindex{} were close to the true peaks and troughs of $\Sinplane$ when those existed at all and that two distinct signal peaks became apparent when $\peakindexm{}\approx 1$. 
At $\alpha=45^{\circ}$ for instance, the second peak appeared at $b\approx 3000$ with $\peakindexm{}\approx 1$ for linear encoding ($c_L\!=\!1$, green dashes) while for planar encoding ($c_L\!=\!0$, pink lines) the transition occurred between $b=\num{5000}$ and $b=\num{10000}$ with \peakindex{} increasing from 0.93 to 1.7. In regimes where $\peakindexm{} > 1$, higher \peakindex{} was associated with sharper signal peaks. Planar encoding signal was higher than signal from prolate encoding with $c_L\!=\!\frac{2}{3}$ (i.e.,$\left|c_L\!-\!\third{}\right| \!=\! \third{} $, continuous pink and green lines), at identical \peakindex{}.\\

% BUTTERFLIES

% PEAKS AND TROUGHS CURVES

\subsection{Robustness of \peakindex{} beyond our toy model}
\begin{figure}
\centering
\includegraphics[width=\textwidth]{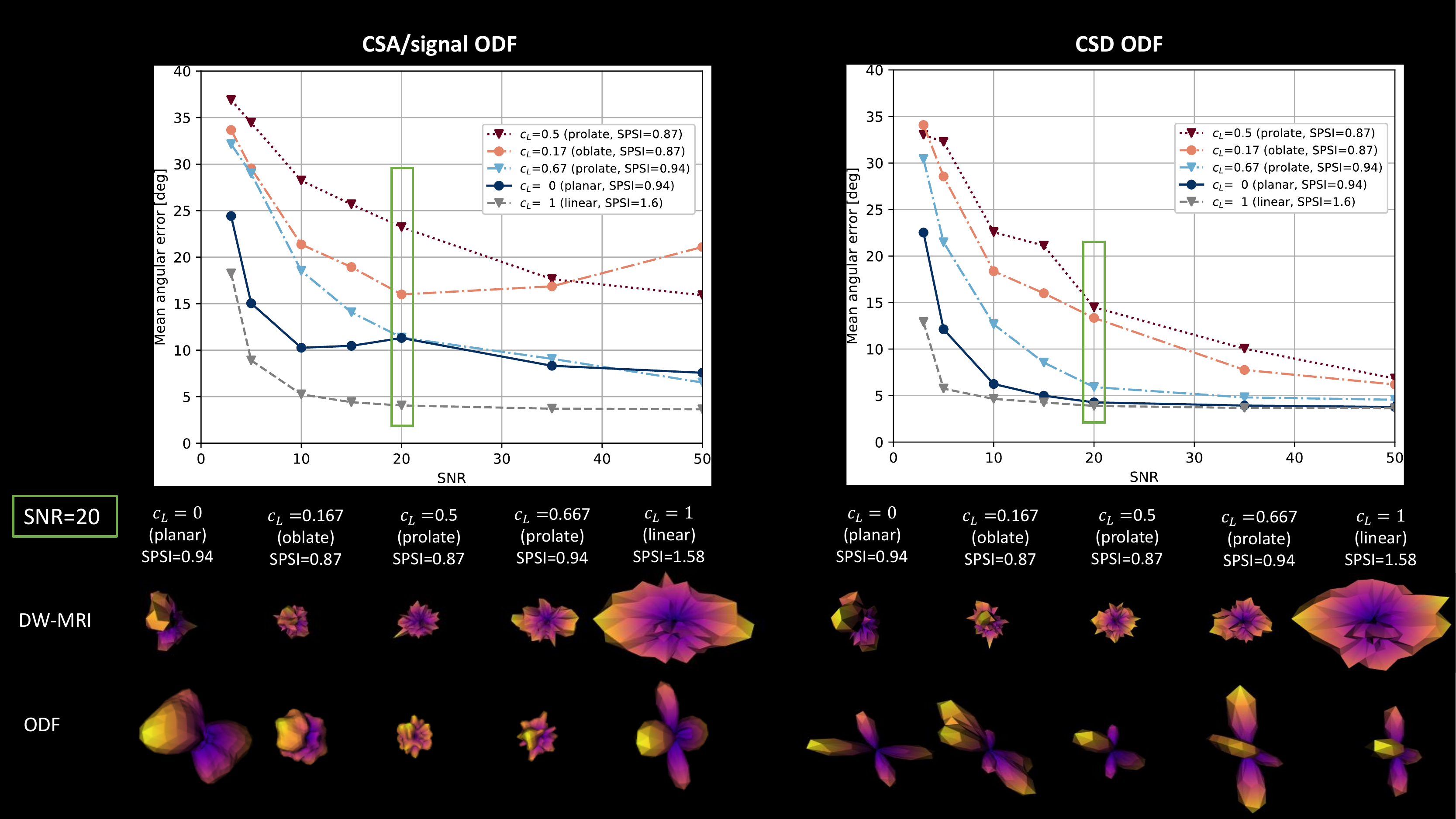}
\caption{\textbf{High \peakindex{} and high signal correlate with better orientation estimation in multi-fascicle, multi-compartment tissue.} Top row: mean angular error (MAE) vs SNR for various types of axisymmetric B-tensor encodings at $b=\SI{3000}{\second\per\milli\meter\squared}$. Bottom row: example of ODF reconstruction from noisy DW-MRI. Left: peaks extracted from the signal ODF~\cite{ozarslan2015rotating} (planar-like encoding) or the CSA ODF~\cite{aganj2010reconstruction} (linear-like encoding). Right: peaks extracted from the CSD ODF~\cite{tournier2007robust} (all encodings).}
\label{fig:odf_rec}
\end{figure}

Even in the context of a more complex multi-fascicle, multi-compartment signal model, lower mean angular error (MAE) in orientation estimation was linked to higher \peakindex{}, i.e. higher $\left|c_L-1/3\right|$, as evidenced in Fig.~\ref{fig:odf_rec}. At equal \peakindex{} or $\left|c_L-1/3\right|$ ($c_L\in\left\{0,\frac{2}{3}\right\}$ and $c_L\in \left\{\frac{1}{6},\frac{1}{2}\right\}$), oblate performed slightly better than prolate encoding. Linear encoding, which maximizes \peakindex{}, consistently outperformed all other encodings across all SNR values.

% Performance seemed mainly driven by $\left|c_L-1/3\right|$. Encodings equidistant from $c_L=1/3$ and thus having similar \peakindex{} had similar results although a slight advantage to prolate was visible, especially with CSD ODF reconstruction. Linear encoding, which achieves $\left|c_L-1/3\right|=2/3$, consistently outperformed all other encodings.
% CSA ODF: with a $8.9^{\circ}$ MAE at SNR=5 steadily decreasing to $3.7^{\circ}$ at SNR=50.
% MAE planar - MAE linear = 6.31414847, 5.49608319, 5.78672983, 4.36009659, 3.15577099 deg at SNR 5, 10, 20, 30, 50

% ODF RECONSTRUCTION

\section{Discussion and conclusion}
\textbf{Limitations.}
Because different gradient waveforms $\v{g}(t)$ (e.g., pulsed and oscillating gradients) may lead to the same $\v{B}$, the B-tensor is a convenient but incomplete representation of a gradient waveform. The zeppelin is also an incomplete picture of diffusion in a fascicle of aligned axons, ignoring among others undulation, orientation dispersion and diffusion restriction. In practice, the actual physics of the gradient waveforms and biological features of tissues likely affect the ability of a sequence to estimate fascicle orientations. Although Fig.~\ref{fig:odf_rec} suggested robustness to deviations from model assumptions, our \peakindex{} could be improved with a more realistic interaction between tissue and sequence via $\eccenD{}\!=\!\left(\lampar \!-\! \lamperp\right)$, by making $\lampar$ and $\lamperp$ functions of, for instance, the sequence's oscillating frequency, diffusion times, gradient separation angle and of the tissue's dispersion, undulation or density.
The proposed index is a signal ratio unaffected by the actual signal intensities, which Eq.~\eqref{eq:ratio_planar_over_linear_same_anisotropy} and \eqref{eq:ratio_planar_over_linear_pure} have shown can be different at identical \peakindex{}. This facilitated rigorous mathematical analysis but may become problematic when the signal approaches the noise floor (e.g., high $b$ and low SNR).
 %However Fig.~\ref{fig:odf_rec} suggests that \peakindex{} meaningfully captured the global performance of a sequence across SNR levels. Considering a signal difference rather than a signal ratio will be considered in future work.
Finally, future analyses should consider non axisymmetric B-tensors (i.e. with a non-zero asymmetry factor as in~\cite{eriksson2015nmr}), possibly via their decomposition into simpler planar, spherical and linear B-tensors, as well as an extension to three-way fascicle crossings.\\
% However, azimuthally symmetric tensors have the obvious advantage of possessing a symmetry axis useful to define acquisition direction.\\

\textbf{\peakindex{} as a tool for sequence design.}
The proposed \peakindex{} (Eq.~\eqref{eq:def_index}) allows encoding parameters ($c_L,b$) to be tuned to target tissue properties such as the expected crossing angle $a$, fascicles' signal fractions $\nu_1$ and microscopic anisotropy $\eccenD{}$. Pathological WM pathways may for instance have low volume and thus low signal contribution $(\nu_2\approx 0)$ or low bundle coherence (low $\eccenD{}$), which may decrease the \peakindex{} at given $c_L,b$ and negatively impact fascicle orientation or ODF estimation. \peakindex{} is a conservative metric as values less than 1 may sometimes be related to in-plane signal already exhibiting two distinct peaks. Ensuring $\peakindexm{}>1$ should thus generally provide a robust safety margin.\\

\textbf{Linear vs planar encoding.}
Linear encoding uniquely maximizes signal contrast via its tensor anisotropy $\left|c_L-\frac{1}{3}\right|$ (equivalently, $\left|b_{\Delta}\right|$~\cite{eriksson2015nmr}), as shown by Eq.~\eqref{eq:def_index}, \eqref{eq:equal_nus_condition_biss}, \eqref{eq:right_angle_cond_maxima} and \eqref{eq:right_angle_cond_minima}. However, Eq.~\eqref{eq:ratio_planar_over_linear_pure} shows that planar always provides higher signal intensities than linear encoding, which might make planar encoding more robust to acquisition noise, ignoring any dependence of the SNR on the type of encoding. Our experiments (Fig.~\ref{fig:odf_rec}) showed that linear systematically outperformed planar encoding, suggesting that the benefits of signal contrast outweigh those of higher signal values. Consistent with Eq.~\eqref{eq:ratio_planar_over_linear_same_anisotropy}, oblate performed better than prolate at fixed \peakindex{} (i.e., identical $\left|c_L-\frac{1}{3}\right|$), possibly owing to their increased robustness to noise. Moderately prolate B-tensors ($c_L\!\in\!\left[\third{},\frac{2}{3}\right]$) therefore seem to offer little benefit for fascicle orientation estimation.

Moreover, a practical drawback of general gradient waveforms compared to linear encoding such as the PGSE is the long durations that they still require to achieve the high b-values~\cite{jensen2018characterizing,topgaard2019diffusion,avram2019measuring,coelho2019resolving,cottaar2020improved} needed to resolve difficult crossings with small angle or a very dominant fascicle (Eq.~\eqref{eq:equal_nus_condition_biss}, \eqref{eq:right_angle_cond_maxima}, \eqref{eq:right_angle_cond_minima}). 
Longer sequences are subject to important T2-decay, which adversely affects SNR. Hybrid protocols combining different types of B-tensor encodings~\cite{lundell2015diffusion}, possibly spread over multiple b-values, may advantageously combine robustness to noise and high signal contrast in fascicle crossings.

\section*{Acknowledgments}
This work was supported by the Swiss National Science Foundation Spark grant number 190297 and has received funding from the European Union’s Horizon 2020 research and innovation programme under the Marie Skłodowska-Curie grant agreement No. 754462.

%
% ---- Bibliography ----
%
% BibTeX users should specify bibliography style 'splncs04'.
% References will then be sorted and formatted in the correct style.
%
 \bibliographystyle{splncs04}
 \bibliography{bibliography}

\appendix

\section{Theory}

\subsection{Single-zeppelin signal for arbitrary orientation under any axisymmetric B-tensor}
\label{app:single_zeppelin_signal}

This section details how Eq.~\eqref{eq:sig_sing} was obtained.
\subsubsection{Rotation of a symmetric tensor}
Any 3-by-3 symmetric matrix $\v{A}$ has real-valued eigenvalues $\Lambda_A=\left\{\lambda_1, \lambda_2, \lambda_3\right\}$ associated to three orthogonal eigenvectors $\uv{u}_1,\uv{u}_2,\uv{u}_3$ and can be diagonalized as
\begin{equation}
\v{A}=
\underbrace{
\left[ \uv{u}_1,\uv{u}_2,\uv{u}_3\right]
}_{\coloneqq \v{R}_{\v{A}}}
 \cdot
\underbrace{
\begin{bmatrix}
\lambda_1 & 0 & 0\\
0 & \lambda_2 & 0\\
0 &0 &\lambda_3
\end{bmatrix}
}_{\coloneqq \v{A}_{z}}
\cdot 
\begin{bmatrix}
\uv{u}_1^\top \\
\uv{u}_2^\top \\
\uv{u}_3^\top
\end{bmatrix}
=
\v{R}_{\v{A}} \v{A}_z \v{R}_{\v{A}}^\top,
\end{equation}
where $\v{R}_{\v{A}}$ is an orthonormal matrix, i.e. satisfying $\v{R}_{\v{A}}^\top\v{R}_{\v{A}}=\v{R}_{\v{A}}\v{R}_{\v{A}}^\top=\v{I}_3$. 
In the axisymmetric case $\lambda_1=\lambda_2$, the eigenvectors $\uv{u}_1$ and $\uv{u}_2$ are defined up to a rotation about $\uv{u}_3$, defined as the symmetry axis. The matrix $\v{R}_{\v{A}}$ is then the rotation matrix mapping  $\uv{u}_3$ onto $\uv{e}_z$.

In general, a rotation matrix $\v{R}_{\uv{u}}$ satisfying $\v{R}_{\uv{u}}\cdot \uv{u}=\uv{e}_z$ is computed as follows, for any unitary vector $\uv{u}=\left[u_x,u_z, u_z\right]^\top\neq \uv{e}_z$, 
\begin{equation}
\v{R}_{\uv{u}}=
\begin{bmatrix}
\frac{(1-u_z)u_y^2}{u_x^2+u_y^2}+u_z, & -\frac{(1-u_z)u_xu_y}{u_x^2+u_y^2}, & \frac{\sqrt{1-u_z^2}u_x}{\sqrt{u_x^2+u_y^2}}\\
-\frac{(1-u_z)u_xy_y}{u_x^2+u_y^2}, &  \frac{(1-u_z)u_x^2}{u_x^2+u_y^2} + u_z, & \frac{\sqrt{1-u_z^2}u_y}{\sqrt{u_x^2+u_y^2}}\\
-\frac{\sqrt{1-u_z^2}u_x}{\sqrt{u_x^2+u_y^2}}, &  -\frac{\sqrt{1-u_z^2}u_y}{\sqrt{u_x^2+u_y^2}}, & u_z
\end{bmatrix}.
\end{equation}
In particular, for a tensor with symmetry axis $\uv{u}=\left[0, \sin\left(\varphi\right), \cos\left(\varphi\right)\right]^\top$ located in the yz-plane, which corresponds to a rotation of $\uv{e}_z$ by an angle $\varphi$ about $\uv{e}_x$ towards the positive y-axis, the rotation matrix $\v{R}_{x}$ can be written as a function of the rotation angle $\varphi$ around $\uv{e}_x$, with $\varphi\in \left[0,2\pi\right]$,
\begin{equation}
\v{R}\left(\varphi\right)=
\begin{bmatrix}
1 & 0 & 0 \\
0 & \cos\left(\varphi\right) & \sin\left(\varphi\right)\\
0 & -\sin\left(\varphi\right) & \cos\left(\varphi\right)
\end{bmatrix}.
\end{equation}

\subsubsection{Rotated diffusion tensor and encoding tensor}
Let $\v{D}$ denote an axisymmetric diffusion tensor with spectrum $\Lambda_D=\left\{\lamperp, \lamperp, \lampar\right\}$ with $0\leq \lamperp<\lampar$. The eigenvector $\uv{u}_D$ associated with $\lampar$ is the symmetry axis of $\v{D}$ and is defined as its orientation. Let $\v{B}$ denote an axisymmetric encoding B-tensor with spectrum $\Lambda_B=\left\{\bperphalf, \bperphalf, \bpar\right\}$, where the eigenvector associated with $\bpar$ (not necessarily the largest eigenvalue) similarly coincides with the symmetry axis of $\v{B}$ and unequivocally defines its orientation, with $\bpar,\bperp\geq 0$.

Without loss of generality, both $\uv{u}_D$ and $\uv{u}_B$ are assumed to lie in the yz-plane an angle $\varphi_D$ and $\varphi_B$ from $\uv{e}_z$ respectively towards the positive y-axis, which leads to
\begin{equation}
\begin{split}
\v{D} &= \v{R}_x\left(\varphi_D\right)
\begin{bmatrix}
\lamperp & 0 & 0 \\
0 & \lamperp & 0 \\
0 &0 & \lampar 
\end{bmatrix}
\v{R}_x\left(\varphi_D\right)^\top \\
&=\begin{bmatrix}
\lamperp  &  0 &  0 \\
0  & \left(\cos^2\left(\varphi_D\right) \lamperp + \sin^2\left(\varphi_D\right) \lampar\right) & \left(\cos\left(\varphi_D\right)\sin\left(\varphi_D\right)\left(\lampar{}-\lamperp{}\right)\right) \\
0 & \left(\cos\left(\varphi_D\right)\sin\left(\varphi_D\right)\left(\lampar{}-\lamperp{}\right) \right) & \left( \sin^2\left(\varphi_D\right) \lamperp + \cos^2\left(\varphi_D\right) \lampar \right)
\end{bmatrix}
\end{split}
\end{equation}
and
\begin{equation}
\begin{split}
\v{B} &= \v{R}_x\left(\varphi_B\right)
\begin{bmatrix}
\bperphalf & 0 & 0 \\
0 & \bperphalf & 0 \\
0 &0 & \bpar 
\end{bmatrix}
\v{R}_x\left(\varphi_B\right)^\top \\
&= \begin{bmatrix}
 \bperphalf  & 0   & 0\\
0  & \left(\cos^2\left(\varphi_B\right) \bperphalf{} + \sin^2\left(\varphi_B\right) \bpar{}\right) & \left(\cos\left(\varphi_B\right)\sin\left(\varphi_B\right)\left(\bpar{}-\bperphalf{}\right)\right) \\
0 & \left(\cos\left(\varphi_B\right)\sin\left(\varphi_B\right)\left(\bpar{}-\bperphalf{}\right) \right) & \left( \sin^2\left(\varphi_B\right) \bperphalf{} + \cos^2\left(\varphi_B\right) \bpar{} \right) 
\end{bmatrix}.
\end{split}
\end{equation}

\subsubsection{Normalized DW-MRI signal}
\begin{figure}
\centering
\includegraphics[scale=0.5, clip=true, trim=0cm 12.8cm 26.0cm 0cm]{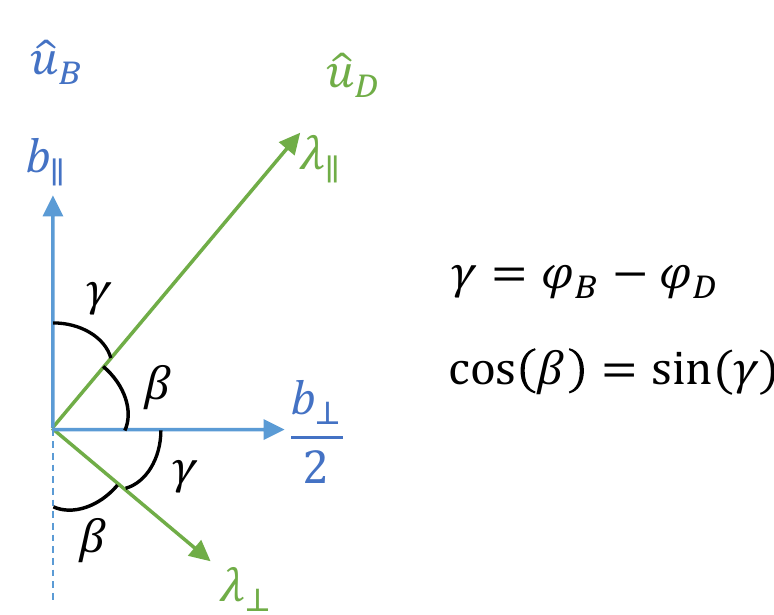}
\caption{\textbf{Geometric interpretation of single-fascicle DW-MRI signal attenuation (Eq.~\eqref{eq:sig_sing}).} 
The interaction between $\bpar{}$ and $\lampar{}$ occurs via $\cos (\gamma )$ 
while that between $\bpar{}$ and $\lamperp{}$ 
via $\cos (\beta )=\sin (\gamma )$ for instance.}
\label{fig:theory_geometric_interp}
\end{figure}

The normalized DW-MRI signal $\Ssing$ arising from a fascicle characterized by a zeppelin $\v{D}$ subject to $\v{B}$ is $\exp\left(-\v{B}:\v{D}\right)$~\cite{neeman1990pulsed}, where $:$ denotes the Frobenius inner product. We compute
\begin{align}
\v{B}:\v{D} =& \bperphalf{}\lamperp{} \nonumber\\
	&+ \cos^2(\varphi_B)\cos^2(\varphi_D) \bperphalf{}\lamperp{} + \sin^2(\varphi_B)\cos^2(\varphi_D)\bpar{}\lamperp{} \nonumber\\
	&+ \cos^2(\varphi_B)\sin^2(\varphi_D) \bperphalf{}\lampar{} + \sin^2(\varphi_B)\sin^2(\varphi_D) \bpar{}\lampar{} \nonumber\\
	&+ \sin^2(\varphi_B)\sin^2(\varphi_D) \bperphalf{}\lamperp{} + \cos^2(\varphi_B)\sin^2(\varphi_D) \bpar{}\lamperp{} \nonumber\\
	&+ \sin^2(\varphi_B)\cos^2(\varphi_D) \bperphalf{}\lampar{} + \cos^2(\varphi_B)\cos^2(\varphi_D) \bpar{}\lampar{}   \nonumber\\
   & + 2 \cos(\varphi_D)\sin(\varphi_D)\cos(\varphi_B)\sin(\varphi_B)\left(\bpar{}-\bperphalf{}\right)\left(\lampar{}-\lamperp{}\right) \nonumber\\
 =& \bperphalf{}\lamperp{}\nonumber \\
  &+ \left(\bperphalf{}\lamperp{}+\bpar{}\lampar{}\right) \left[\left(\cos(\varphi_B)\cos(\varphi_D)\right)^2 + \left(\sin(\varphi_B)\sin(\varphi_D)\right)^2 \right] \nonumber\\
  &+ \left(\bpar{}\lamperp{}+\bperphalf{}\lampar{} \right)\left[\left(\sin(\varphi_B)\cos(\varphi_D)\right)^2 + \left(\cos(\varphi_B)\sin(\varphi_D)\right)^2\right] \nonumber\\
  &+ 2\left(\bpar{}-\bperphalf{}\right)\left(\lampar{}-\lamperp{}\right) \cos(\varphi_D)\sin(\varphi_D)\cos(\varphi_B)\sin(\varphi_B). \label{eq:BD_start}
\end{align}
%  &+ \bperphalf{}\lampar{}\left[\left(\cos(\varphi_B)\sin(\varphi_D)\right)^2 + \left(\sin(\varphi_B)\cos(\varphi_D)\right)^2\right] \\
%  &+ \bpar{}\lampar{} \left[\left(\sin(\varphi_B)\sin(\varphi_D)\right)^2 + \left(\cos(\varphi_B)\cos(\varphi_D)\right)^2 \right]\\
Using the following relationships (Simpson's formulas)
\begin{subequations}
\begin{align}
\cos a \cos b = \frac{\cos(a+b) + \cos(a-b)}{2} \label{eq:simpson_cpc}\\
\sin a \sin b = \frac{\cos(a-b) - \cos(a+b)}{2} \label{eq:simpson_cmc}\\
\cos a \sin b = \frac{\sin(a+b) - \sin(a-b)}{2} \nonumber\\
\sin a \cos b = \frac{\sin(a+b) + \sin(a-b)}{2}, \nonumber
\end{align}
\end{subequations}
Eq.~\eqref{eq:BD_start} becomes
\begin{align*}
\v{B}:\v{D} =&  \bperphalf{}\lamperp{} \\
	&+ \left(\bperphalf{}\lamperp{}+\bpar{}\lampar{}\right) \left(\frac{\cos^2(\varphi_B+\varphi_D) + \cos^2(\varphi_B-\varphi_D) }{2}\right) \\
	&+ \left(\bpar{}\lamperp{}+\bperphalf{}\lampar{}\right) \left(\frac{\sin^2(\varphi_B+\varphi_D) + \sin^2(\varphi_B-\varphi_D)}{2} \right)\\
	&+ \underbrace{\left( \lampar{}-\lamperp{}\right)\left(\bpar{}-\bperphalf{}\right)}_{=\left(\bperphalf{}\lamperp{}+\bpar{}\lampar{}\right)-\left(\bpar{}\lamperp{}+\bperphalf{}\lampar{}\right)} \left(\frac{\cos^2(\varphi_B-\varphi_D) - \cos^2(\varphi_B+\varphi_D)}{2} \right)\\
 =&  \bperphalf{}\lamperp{} \\
  &+ \left(\bperphalf{}\lamperp{}+\bpar{}\lampar{}\right) \cos^2\left(\varphi_B-\varphi_D\right) \\
  &+ \left(\bpar{}\lamperp{}+\bperphalf{}\lampar{}\right) \left( \frac{\sin^2(\varphi_B+\varphi_D)+\cos^2(\varphi_B+\varphi_D) - \cos^2(\varphi_B-\varphi_D) + \sin^2(\varphi_B+\varphi_D)}{2} \right)\\
 =& \bperphalf{}\lamperp{}
    + \left(\bperphalf{}\lamperp{}+\bpar{}\lampar{}\right) \cos^2\left(\varphi_B-\varphi_D\right)
    + \left(\bpar{}\lamperp{}+\bperphalf{}\lampar{}\right) \sin^2(\varphi_B-\varphi_D),
\end{align*}
which leads to the final expression for $\exp\left(-\v{B}:\v{D}\right)$ in Eq.~\eqref{eq:sig_sing}. Figure~\ref{fig:theory_geometric_interp} proposes an informal geometric interpretation for this formula.

\FloatBarrier

\subsection{Approximate max-to-min ratio of in-plane signal}
\label{app:max_to_min_ratio}
This section details how to obtain Eq.~\eqref{eq:def_index}, an explicit formula for the proposed \peakindex{}, from its definition in Eq.~\eqref{eq:def_index_via_ratio}.\\

In the planar-like case $c_L\leq \third{}$, recalling that $\nu_1+\nu_2=1$ and using the definition of $\Ssing{}$ in Eq.~\eqref{eq:sig_sing},
\begin{align}
\frac{\Sinplane\left(\alpha\right)}{\Sinplane\left(\alpha/2\right)}
 =& \frac{\nu_1 \Ssing(\alpha) + \nu_2 \Ssing(0)}
         {\nu_1\Ssing\left(\frac{\alpha}{2}\right)
         + \nu_2\Ssing\left(\frac{\alpha}{2}\right)}
         = \frac{ \nu_1 \Ssing(\alpha) + \nu_2 \Ssing(0)}{\Ssing\left(\frac{\alpha}{2}\right)} \nonumber\\
 =& \nu_1 \frac{ \exp\left\{ -\bperphalf{}\lamperp{}
    				- \left(\bperphalf{}\lamperp{}+\bpar{}\lampar{}\right) \cos^2\left(\alpha\right)
    				- \left(\bpar{}\lamperp{}+\bperphalf{}\lampar{}\right) \sin^2(\alpha)
    			\right\}
 	}
 	{ \exp\left\{ -\bperphalf{}\lamperp{}
    			- \left(\bperphalf{}\lamperp{}+\bpar{}\lampar{}\right) \cos^2\left(\frac{\alpha}{2}\right)
    			- \left(\bpar{}\lamperp{}+\bperphalf{}\lampar{}\right) \sin^2\left(\frac{\alpha}{2}\right)
    		\right\}
 	} \nonumber\\
 	&+\nu_2 \frac{ \exp\left\{ -\bperphalf{}\lamperp{} - \left(\bperphalf{}\lamperp{}+\bpar{}\lampar{}\right) \right\}}
 	    {\exp\left\{-\bperphalf{}\lamperp{}
    			- \left(\bperphalf{}\lamperp{}+\bpar{}\lampar{}\right) \cos^2\left(\frac{\alpha}{2}\right)
    			- \left(\bpar{}\lamperp{}+\bperphalf{}\lampar{}\right) \sin^2\left(\frac{\alpha}{2}\right)
    		\right\}
 	    } \nonumber\\
 =& \nu_1 \exp\left\{ \left(\bperphalf{}\lamperp{}+\bpar{}\lampar{}\right) 
 					 \left( \cos^2\left(\frac{\alpha}{2}\right) - \cos^2\left(\alpha\right) \right) 
 					 + \left(\bpar{}\lamperp{}+\bperphalf{}\lampar{}\right)
 					   \left( \sin^2\left(\frac{\alpha}{2}\right) - \sin^2(\alpha)\right)
              \right\} \nonumber \\
    &+\nu_2 \exp\left\{ \left(\bperphalf{}\lamperp{}+\bpar{}\lampar{}\right)
    				    \left(\cos^2\left(\frac{\alpha}{2}\right)-1\right)
    				   + \left(\bpar{}\lamperp{}+\bperphalf{}\lampar{}\right)\sin^2\left(\frac{\alpha}{2}\right)    				     
    			\right\}. \label{eq:theory_ratio}
\end{align}
Using Eq.~\eqref{eq:simpson_cpc} and \eqref{eq:simpson_cmc}, the following relationships hold
\begin{align*}
\cos^2a - \cos^22a
	=& \left( \cos a + \cos 2a\right) \left(\cos a - \cos 2a\right) &\\
	=& \left(2 \cos\frac{3a}{2}\cos\frac{a}{2}\right) \left(-2 \sin\frac{3a}{2} \sin\left(-\frac{a}{2}\right)\right)  \\
	=& 2 \sin\frac{3a}{2} \cos\frac{3a}{2} 2 \sin\frac{a}{2} \cos\frac{a}{2}\\
	=& \sin 3a \sin a
  	& &= - \left( \sin^2 a - \sin^2 2a \right),
\end{align*}
and Eq.~\eqref{eq:theory_ratio} can be rewritten as
\begin{align*}
\frac{\Sinplane\left(\alpha\right)}{\Sinplane\left(\alpha/2\right)}
 =& \nu_1 \exp\left\{ \sin\left(\frac{3\alpha}{2}\right) \sin\left(\frac{\alpha}{2}\right) 
 					  \left(\bperphalf{}\lamperp{}+\bpar{}\lampar{}-\bpar{}\lamperp{}-\bperphalf{}\lampar{}\right)
 			\right\} \nonumber \\
 	&+ \nu_2 \exp\left\{ -\sin^2\left(\frac{\alpha}{2}\right)
 	                     \left(\bperphalf{}\lamperp{}+\bpar{}\lampar{}-\bpar{}\lamperp{}-\bperphalf{}\lampar{}\right)
 			\right\}, \nonumber \\
 =& \nu_1 \exp\left\{\frac{3}{2} \sin\left(\frac{3\alpha}{2}\right) \sin\left(\frac{\alpha}{2}\right)
 				      \clmthird{}b\eccenD{}
              \right\}
     + \nu_2 \exp\left\{ -\frac{3}{2}\sin^2\left(\frac{\alpha}{2}\right)
     					\clmthird{}b\eccenD{}
		     \right\},
\end{align*}
where the following was used
\begin{align*}
\left(\bperphalf{}\lamperp{}+\bpar{}\lampar{}-\bpar{}\lamperp{}-\bperphalf{}\lampar{}\right)
 = \left(\bpar{}-\bperphalf{}\right)\left(\lampar{}-\lamperp{}\right)= \frac{3}{2}\clmthird{}b \underbrace{\left(\lampar{}-\lamperp{}\right)}_{\eccenD{}}.
\end{align*}

In the linear-like case $c_L>\third{}$, the other branch of the definition in Eq.~\eqref{eq:def_index_via_ratio} is used, which
 has the effect of replacing every instance of $\cos(\cdot)$ by $\sin(\cdot)$ and vice versa in Eq.~\eqref{eq:theory_ratio}, and eventually leads to
\begin{align*}
&\frac{\Sinplane\left(\alpha-\pi/2\right)}{\Sinplane\left(\alpha/2 -\pi /2\right)}\\
 &\quad = \nu_1 \exp\left\{-\frac{3}{2} \sin\left(\frac{3\alpha}{2}\right) \sin\left(\frac{\alpha}{2}\right)
 				      \clmthird{}b\eccenD{}
              \right\}
     + \nu_2 \exp\left\{\frac{3}{2}\sin^2\left(\frac{\alpha}{2}\right)
     					\clmthird{}b\eccenD{}
		     \right\}.
\end{align*}
The linear-like and planar-like cases can be compactly summarized using the $\left|c_L-\third{}\right|$ notation, which is the final form of Eq.~\eqref{eq:def_index}.

%% DERIVATIVE !
%Its derivative with respect to $\varphi_B$
%\begin{equation}
%\begin{split}
%\frac{\partial S\left(\varphi_B\right)}{\partial \varphi_B} & = \\
%	\frac{3}{2} & \clmthird{} b \left(\lampar-\lamperp\right) 
%			\left[\nu_1 \sin\left(2\varphi_B\right) \Ssing\left(\varphi_B\right) 
%				+\nu_2 \sin\left(2\left(\alpha-\varphi_B\right)\right) \Ssing\left(\alpha-\varphi_B\right)
%			\right]
%\end{split}
%\label{eq:derivative_inplane}
%\end{equation}

\subsection{Ratios of planar to linear signal}
\label{app:all_ratios}
This section provides additional ratios of planar to linear signal in a total of 8 different cases recapped in Tab.~\ref{tab:ratios_planar_over_linear}: in single-fascicle or in crossing-fascicle voxels, with equal B-tensor anisotropy $\left|c_L-\third{}\right|$ or with extremal anisotropies $c_L=0$ vs $c_L=1$, with or without a $\frac{\pi}{2}$ shift in one of the two signals. 
The case of crossing-fascicle voxels with equal B-tensor anisotropy and a $\frac{\pi}{2}$ shift in the denominator corresponds to Eq.~\eqref{eq:ratio_planar_over_linear_same_anisotropy} in the text while Eq.~\eqref{eq:ratio_planar_over_linear_pure} corresponds to the same scenario but with extreme anisotropies, i.e. purely linear over purely planar.

When the $\frac{\pi}{2}$ shift is applied, the planar or oblate signal can be shown to be (strictly) greater than the linear or prolate signal. The proof is provided in the next section for the most complex case, i.e. crossing fascicles and extreme B-tensor anisotropies (Eq.~\eqref{eq:ratio_planar_over_linear_pure} in the text).

\begin{table}[h!]
\centering
\caption{\textbf{Ratios of planar to linear signals.} For single fascicles, $S$ refers to $\Ssing{}$ and $\varphi$ to the angle between the orientations of the B-tensor $\uv{u}_B$ and of the unique zeppelin $\uv{u}_D$. For crossing fascicles, $S$ refers to $\Sinplane{}$ and $\varphi$ to the azimuthal angle of $\uv{u}_B$ computed from $\uv{u}_1$ in the plane defined by $\uv{u}_1$ and $\uv{u}_2$. Inequalities are understood $\forall \varphi$ and given that $b,\eccenD{}, \Delta_L>0$ and $\alpha\in\left[0,\frac{\pi}{2}\right]$. We defined $K_1\coloneqq K_1(b,\eccenD{},\varphi)=\frac{b}{2}\eccenD{}\left(3\cos^2(\varphi)-1\right)$ and $K_2\coloneqq K_2(b,\eccenD{},\varphi,\alpha)=\frac{b}{2}\eccenD{}\sin(\alpha)\sin(2\varphi-\alpha)$.}
\vskip 0.5cm 
\begin{tabular}{l@{\hskip 0.5cm}c@{\hskip 0.6cm}c}
   & Single fascicle    & Crossing fascicles \\ % column headers
                    \hline
\begin{tabular}{@{}c@{}}Symmetric, $\frac{\pi}{2}$ shift \\
$\frac{S\left(\varphi;c_L=\frac{1}{3}\!-\!\Delta_L \right)}{S\left(\varphi\!-\!\frac{\pi}{2};c_L=\third{}\!+\!\Delta_L \right)}$
\end{tabular} % left column
    & $\expbrace{\frac{b}{2}\Delta_L \eccenD{}} > 1 $     
    & $\expbrace{\frac{b}{2}\Delta_L \eccenD{} } > 1 $  \\ % main table
\begin{tabular}{@{}c@{}}Symmetric, no $\frac{\pi}{2}$ shift
\\$\frac{S\left(\varphi;c_L=\third{}\!-\!\Delta_L \right)}{S\left(\varphi;c_L=\third{}\!+\!\Delta_L \right)}$
\end{tabular}  
	& $\expbrace{2\Delta_L K_1}$ 
	& $\expbrace{2\Delta_L K_1} \left[\frac{\nu_1+\nu_2\expbrace{3\Delta_L K_2}}{\nu_1+\nu_2\expbrace{-3\Delta_L K_2}}\right]$         \\ % main table
\begin{tabular}{@{}c@{}}Extremal, $\frac{\pi}{2}$ shift \\
$\frac{S\left(\varphi;c_L=0\right)}{S\left(\varphi\!-\!\frac{\pi}{2};c_L=1 \right)}$
\end{tabular} 
	& $\expbrace{\frac{b}{2}\eccenD{}\sin^2(\varphi)} \geq 1 $            
	& $\expbrace{\frac{b}{2}\eccenD{}\sin^2(\varphi)} \left[\frac{\nu_1+\nu_2\expbrace{K_2}}{\nu_1 + \nu_2\expbrace{2K_2}}\right] >1 $    \\ % main table
\begin{tabular}{@{}c@{}}Extremal, no $\frac{\pi}{2}$ shift \\
$\frac{S\left(\varphi;c_L=0\right)}{S\left(\varphi;c_L=1\right)}$
\end{tabular}  
	& $\expbrace{K_1}$ 
	& $\expbrace{K_1} \left[\frac{\nu_1+\nu_2\expbrace{K_2}}{\nu_1+\nu_2\expbrace{-2K_2}}\right]$       % main table
\end{tabular}
\label{tab:ratios_planar_over_linear}
\end{table}
\FloatBarrier

\subsection{Crossing-fascicle in-plane planar always greater than shifted in-plane linear signal}
\label{app:proof_difficult_ratio}

This section proves that the ratio of planar to ($90^{\circ}$-shifted) linear in-plane signal is always greater 1 in Eq.~\eqref{eq:ratio_planar_over_linear_pure}. Multiplying the numerator and the denominator by $\exp\left\{ -\sin\left(\alpha\right) \sin\left(2\varphi_B\!-\!\alpha\right) b     \eccenD{} \right\} $, Eq.~\eqref{eq:ratio_planar_over_linear_pure} becomes
\begin{align}
&\frac{\Sinplane{}\left(\varphi_B;c_L\!=\!0 \right)}{\Sinplane{}\left(\varphi_B\!-\!\frac{\pi}{2};c_L\!=\!1 \right)} \nonumber\\
 &=
\exp\left\{\sin^2\left(\varphi_B\right) \frac{b}{2}\eccenD{} \right\}   \cdot
 \left[\frac{\nu_1 \exp\left\{ -\sin\left(\alpha\right) \sin\left(2\varphi_B\!-\!\alpha\right) b  \eccenD{} \right\} 
             + \nu_2 \exp\left\{ -\sin\left(\alpha\right) \sin\left(2\varphi_B\!-\!\alpha\right)\frac{b}{2}\eccenD{} \right\} }
            {\nu_1\exp\left\{ -\sin\left(\alpha\right) \sin\left(2\varphi_B\!-\!\alpha\right)b \eccenD{} \right\} 
            +\nu_2 }
 \right] \nonumber\\
 &= \frac{K_1\nu_1 + K_2 \nu_2}{K_3 \nu_1 + \nu_2}, \label{eq:theory_ratio_linear_planar_pure}
\end{align}
where 
\begin{align*}
K_1 &\coloneqq \exp\left\{\sin^2\left(\varphi_B\right) \frac{b}{2}\eccenD{} \right\} 
			    \exp\left\{ -\sin\left(\alpha\right) \sin\left(2\varphi_B\!-\!\alpha\right) b  \eccenD{} \right\}\\
K_2 &\coloneqq  \exp\left\{\sin^2\left(\varphi_B\right) \frac{b}{2}\eccenD{} \right\}
				\exp\left\{ -\sin\left(\alpha\right) \sin\left(2\varphi_B\!-\!\alpha\right)\frac{b}{2}\eccenD{} \right\} \\
K_3 &\coloneqq  \exp\left\{ -\sin\left(\alpha\right) \sin\left(2\varphi_B\!-\!\alpha\right) b \eccenD{} \right\}.\\
\end{align*}
Since $b\eccenD{}>0$, we have $K_1=K_3$ for $\varphi_B=k\pi \left(k\in\mathbb{Z}\right)$ since then $\exp\left\{\sin^2\left(\varphi_B\right) \frac{b}{2}\eccenD{} \right\} = 1 $, and $K_1>K_3$ for all other values of $\varphi_B$. It therefore remains to prove that $K_2 \geq 1$ and that $K_2=1$ never occurs when $K_1=K_3$ to have $K_1\nu_1 + K_2 \nu_2 > K_3 \nu_1 + \nu_2 $ in Eq.~\eqref{eq:theory_ratio_linear_planar_pure}, since $\nu_2>0$.

Simpson's identity for sines (Eq.~\eqref{eq:simpson_cmc}) and Carnot's formula yield
\begin{align*}
\sin\left(\alpha\right) \sin\left(2\varphi_B\!-\!\alpha\right) &= \frac{\cos(2\alpha-2\varphi_B) - \cos(2\varphi_B)}{2} \\
\sin^2(\varphi_B) &= \frac{1-\cos(2\varphi_B)}{2},
\end{align*}
allowing $K_2$ to be rewritten as
\begin{align*}
K_2 &= \exp\left\{  \frac{1-\cos(2\varphi_B)}{2} \frac{b}{2}\eccenD{} \right\} 
        \exp\left\{ - \frac{\cos(2\alpha-2\varphi_B) - \cos(2\varphi_B)}{2} \frac{b}{2}\eccenD{} \right\} \\
     &= \exp\left\{ \frac{1}{2}
     				\underbrace{\left(1-\cos(2\alpha-2\varphi_B)\right)}_{\geq 0}
     				\frac{b}{2}\eccenD{} \right\} \geq 1,\quad \forall  \varphi_B.
\end{align*}
Strict equality $K_2=1$ only occurs for $\varphi_B=\alpha + \frac{\pi}{4} + k\pi \left( k\in\mathbb{Z} \right)$, which never coincide with the values $\varphi_B=k\pi \left(k\in\mathbb{Z}\right)$ leading to $K_1=K_3$ for a crossing angle $\alpha\in\left[0,\frac{\pi}{2}\right]$. Consequently, Eq.~\eqref{eq:theory_ratio_linear_planar_pure} is always $>1$ and the ratio in Eq.~\eqref{eq:ratio_planar_over_linear_pure} is thus also $> 1$.

\end{document}